\documentstyle[aps,twocolumn,graphicx]{revtex}

\begin{document}
%
\title{Impact on SUSY-Breaking Models of \\
the $R$-Parity Violating Squark Interpretation of the HERA Anomaly}
\author{
Kingman Cheung$^1$, Duane A. Dicus$^1$, and B. Dutta$^2$ \\
$^1${\it Center for Particle Physics, University of Texas, Austin, TX 78712}\\
$^2${\it Institute of Theoretical Science, University of Oregon, Eugene, 
OR 97403}
}
\maketitle

\begin{abstract}  
The explanation of a 200 GeV $R$-parity violating squark to the HERA
high-$Q^2$ anomaly would have an important impact on
supersymmetry-breaking models.  Here we show that a squark
of mass around 200 GeV is disallowed in the minimal
gauge-mediated SUSY-breaking models in the parameter space constrained
by the radiative electroweak-symmetry breaking and the experimental
lower limits on supersymmetric particles, even after including
the large $R$-parity violating couplings.  Supergravity motivated models can,
on the other hand, give rise to these scalar quarks in a wide range of
parameter space.
\end{abstract}

\thispagestyle{empty}

\section{Introduction}
The mechanism of supersymmetry (SUSY) breaking and how it is
communicated to the observable sector is still an unsolved puzzle.  
It has been assumed that the SUSY
is broken in a hidden sector at a scale of $\sim 10^{11}$ GeV and the 
SUSY breaking is communicated to the visible sector via the gravitational
interaction. Recently, another class of models has been proposed
where SUSY is broken in a hidden sector at a scale around $10^{5}$ GeV and 
the SUSY breaking is communicated to the visible sector via the standard model
gauge interactions \cite{GMSB}. Now the question is how to
distinguish between these two scenarios by experiments.
One method is the direct search for their signatures in collider experiments. 
So far all direct searches for SUSY particles at current colliders like
the Fermilab Tevatron, LEP, and HERA have been negative.
Recent results from HERA \cite{H1,zeus,lp} on deep
inelastic scattering have showed an excess of
events in the high-$Q^2$ and large $x,y$ region;
specifically H1 showed an impressive
enhancement in a single $x$ bin corresponding to $M=\sqrt{sx}\simeq
200$ GeV, 7 events were observed where only one was expected.  Further
data has given one more event \cite{lp} so there are now 8 events
where $1.5$ events is expected.

Although still more data is needed to
confirm this excess in the cross section, many attempts have already been
made to explain it.  These include $eeqq$ contact interactions
\cite{contact}, leptoquark production \cite{alta,lq}, and $R$-parity
violating (RPV) squark production \cite{squark}.  It is possible to
construct models of contact interactions, which satisfy the known
constraints such as atomic parity violation, LEP data, and low energy
electron-nucleon and neutrino-nucleon scattering data 
\cite{bchz}, but the effect is very small.  
The leptoquark explanation also runs into trouble
because the latest CDF \cite{LQ-cdf} and D0 \cite{LQ-d0} bounds rule
out the mass of the first generation leptoquark up to 213 and 225 GeV,
respectively, at a 95\% CL assuming the leptoquark decays entirely
into $eq$.  On the other hand, the RPV squark remains a viable solution.
The squark can either
be the left-handed scalar charm $\tilde{c}_L$ or the lighter mass
eigenstate of the scalar top $\tilde{t}_1$ with a mass around 200 GeV.
Such a 200 GeV squark has an important impact on SUSY-breaking
models.

In this paper, we point out that it is almost impossible to generate
such light squarks in gauge-mediated SUSY-breaking models
even if we include large RPV couplings.
We show explicitly in the minimal gauge-mediated models
that squark masses are excluded up to 300 GeV in the
parameter space allowed by the experimental
mass limits of the Higgs bosons, the chargino, and the scalar tau.
On the other hand, in the 
supergravity-motivated models it is possible to generate squarks of
mass around 200 GeV.

\section{The $R$-parity Violating Squark Solution}

The $R$-parity
violation is introduced in the superpotential via additional terms:
\begin{equation} 
{\cal W}_R= 
\lambda_{ijk} L_i L_j E^c_k + \lambda'_{ijk} L_i Q_j D_k^c  +
\lambda^{''}_{ijk}U^c_i D^c_j D^c_k
\end{equation}  
where $L_i,E^c,Q,U^c,D^c$ denote the superfields
and $i,j,k$ are generation indices.  Here we already assume
the absence of the $L_i H_u$ term and $\lambda$ and
$\lambda^{''}$ to be zero because they are not relevant for the HERA
high-$Q^2$ events and zero $\lambda^{''}$'s can avoid rapid proton
decay.

The process relevant to the HERA events
is $e^+ d \to \tilde{c}_L (\tilde{t}_L) \to e^+ d$.
The weak eigenstates
$\tilde{t}_L$ and $\tilde{t}_R$ of the stop mix to form the mass
eigenstates $\tilde{t}_1$ and $\tilde{t}_2$.  The values of
$\lambda'_{121}$ and $\lambda'_{131}$ needed to explain the large
cross sections at HERA are $(0.03-0.04)/\sqrt{B}$ \cite{alta,squark},
where $B$ is the branching ratio for the squark to decay into $e^+
d$. To satisfy the constraints from atomic parity violation and from
the leptoquark search at the Tevatron the branching ratio $B$ must be
within the range $0.3-0.5 \alt B \alt 0.75$ \cite{alta,LQ-cdf,LQ-d0},
which implies $0.03 \alt\lambda'_{121},\lambda'_{131} \alt 0.07$. The
production of $\tilde{u}_L$ cannot explain the anomaly
because the coefficient $\lambda'_{111}$ is tightly constrained by
neutrinoless double beta decay \cite{hirsh}.  The production via sea
partons also requires large $\lambda'$, which are either
already ruled out or close to the allowed
limits \cite{squark}.  Thus, the most likely explanation within SUSY
is the production of $\tilde{c}_L$ or
$\tilde{t}_1$ with $\lambda'_{121}$ or $\lambda'_{131}$ of order
$0.03-0.07$.   These coefficients
$\lambda'_{121}$ and $\lambda'_{131}$ are sufficiently small that the
sparticle spectrum is not affected appreciably by their
presence.  However, there are still some other $\lambda'_{ijk}$ that
are neither constrained by present experiments nor necessarily small
due to symmetry; in particular $\lambda'_{233}$ and $\lambda'_{333}$
\cite{carlo,vern3}. 

The relevant RPV terms in our RGE analysis correspond to 
$\lambda'_{233}$ and $\lambda'_{333}$ in the superpotential ${\cal W}_R$. The
corresponding trilinear terms are $C'_{ijk} \tilde{L_i} 
\tilde{Q_j} \tilde{D_k}$ with $ijk=233,333$.   For example, the RGE for
$M_{Q_L}^2$ is given by \cite{carlo,vern2}
\begin{eqnarray}
\frac{d M^2_{Q_L}}{dt} &=& \frac{2}{16\pi^2} \biggr[ -\frac{1}{15} g_1^2 M_1^2
-3 g_2^2 M_2^2 - \frac{16}{3}g_3^2 M_3^2 + h_t^2 X_t + h_b^2 X_b \nonumber\\ &&
+ {C'_{233}}^2 + {C'_{333}}^2 + {\lambda'_{233}}^2 M^2_{L_2 L_2} 
+{\lambda'_{333}}^2 M^2_{L_3 L_3}  + \lambda'_{233}\lambda'_{333}  (M^2_{L_2
L_3} +M^2_{L_3 L_2} ) \nonumber \\ && + ({\lambda'_{233}}^2 +{\lambda'_{333}}^2
)( M^2_{Q_L} + M^2_{b_R}) 
\biggr ] \label{QL}\;,
\end{eqnarray} 
where the notation can be found in Ref. \cite{carlo,vern2}. 
The first line of
Eq.  (\ref{QL}) contains the $R$-parity conserving contributions and
the rest are $R$-parity violating.
Although $\lambda'_{233}$ and $\lambda'_{333}$ are not constrained by
existing experiments, we restrict them by requiring all Yukawa couplings  and
these $\lambda'$ be perturbative up to the GUT scale. 
We find that in the range
$\tan \beta=2-50$, $\lambda'_{233}$ and $\lambda'_{333}$ are required
to be less than $0.5-0.7$ at the weak scale in order to keep all
Yukawa couplings perturbative.  
\footnote{It was pointed out in Ref. \cite{carlo} that a small neutrino
mass is generated through the vev of the sneutrino by the renormalization 
group equations.  The neutrino mass then constrains severely  the
$\lambda'$.  On the other hand, it was pointed out in Ref. \cite{dreiner}
that there are ambiguities on these bounds due to rotations in the 
($L_i, H_u$) 
space \cite{davidson} and the effect can actually be suppressed by 
a dynamical alignment mechanism \cite{polonsky} or a horizontal symmetry.  We
therefore do not consider this neutrino mass constraint in picking the
values for $\lambda'$.}

\section{The Minimal Gauge-Mediated Models}


In the minimal gauge-mediated model, the standard
model gauge interactions communicate the SUSY breaking to the visible
sector and give masses to the gauginos and the scalars:
\begin{equation}
\label{boundary} 
M_i=n\, g\left(\frac{\Lambda}{M} \right ) \;
\frac{\alpha_i(M)}{4\pi}\, \Lambda \;, \qquad  m_0^2 = 2n
f\left(\frac{\Lambda}{M} \right ) \; 
\Lambda^2 \sum_{i=1}^3 k_i \left( \frac{\alpha_i(M)}{4\pi} \right)^2\;,
\end{equation} 
where $\Lambda$ is the SUSY breaking scale and $M$ is the scale at which
the soft masses are introduced. The sum is
over SU(3)$\times$SU(2)$_L \times$U(1)$_Y$, with $k_1=
\frac{3}{5}(\frac{Y}{2})^2$, $k_2=\frac{3}{4}$ for SU(2)$_L$ doublets
and zero for singlets, and $k_3=\frac{4}{3}$ for color triplets and
zero for color singlets.  $n$ is the number of multiplets in the
messenger sector, and $g(\Lambda/M)$ and $f(\Lambda/M)$ \cite{threshold} 
are the messenger-scale threshold functions.
The soft SUSY-breaking parameters $A$, $C'_{233}$, and $C'_{333}$
are set to zero at the scale $M$ because they are induced only by
higher loops.  

The input parameters are $M$, $\Lambda$, and $\tan\beta=v_2/v_1$,
where $v_1$ and $v_2$ are the vacuum expectation values of the two
Higgs doublets.  We vary the minimal model by adding
more multiplets to the messenger sector, represented by $n>1$.
The procedures for running the RGE is as follows: 
(i) we use the inputs $m_t^{\rm phy}=175$ GeV which is related to
$m_t(m_t)$ by $m_t^{\rm phy}= m_t(m_t) (1+ 4\alpha_s/3\pi )$, 
$m_b(m_b)=4.25$ GeV, $m_\tau(m_\tau)=1.784$ GeV,
$\alpha_{\rm em}(M_Z)=1/128.9$, $\sin^2\theta_{\rm w}=0.23165$, and
$\alpha_s(M_Z)=0.118$.   We chose $M_{\rm weak}=m_t(m_t)$.
The values for $\lambda'_{233}$ and $\lambda'_{333}$ are
chosen between 0.0 and 0.5 at $M_{\rm weak}$; (ii) 
we evolve all the gauge, Yukawa, and RPV couplings from 
$M_{\rm weak}$ to the scale $M$ using the SUSY RGE \cite{carlo,vern2}.  
At the scale $M$ we calculate the
gaugino and scalar masses using Eq. (\ref{boundary});
(iii) we evolve all the soft SUSY parameters except $B$ and $\mu$ from $M$
down to $M_{\rm weak}$.  Then we use the full 1-loop effective
potential and by minimization solve for $\mu$ and $B$.  
The $\mu$ parameter is determined up to a sign, and the CLEO data on the 
inclusive decay $b\rightarrow s\gamma$ prefers $\mu<0$ \cite{ddo}.
In order to maintain perturbative unification we
consider only $n\le 4$.  In the absence of late inflation cosmological
constraints put an upper bound on the gravitino mass of about $10^4$
eV \cite{pp}, which restricts $M/\Lambda=1.1-10^4$.

For the supergravity motivated models we assume the universal boundary
conditions at the GUT scale, i.e, a common scalar mass ($m_0$), a
common gaugino mass $(m_{1/2})$, and a common trilinear coupling ($A$). 
We run all the soft parameters from the GUT scale down to the weak scale.

\section{Results}

\begin{figure}[t]
\leavevmode
\begin{center}
\includegraphics[height=2.4in]{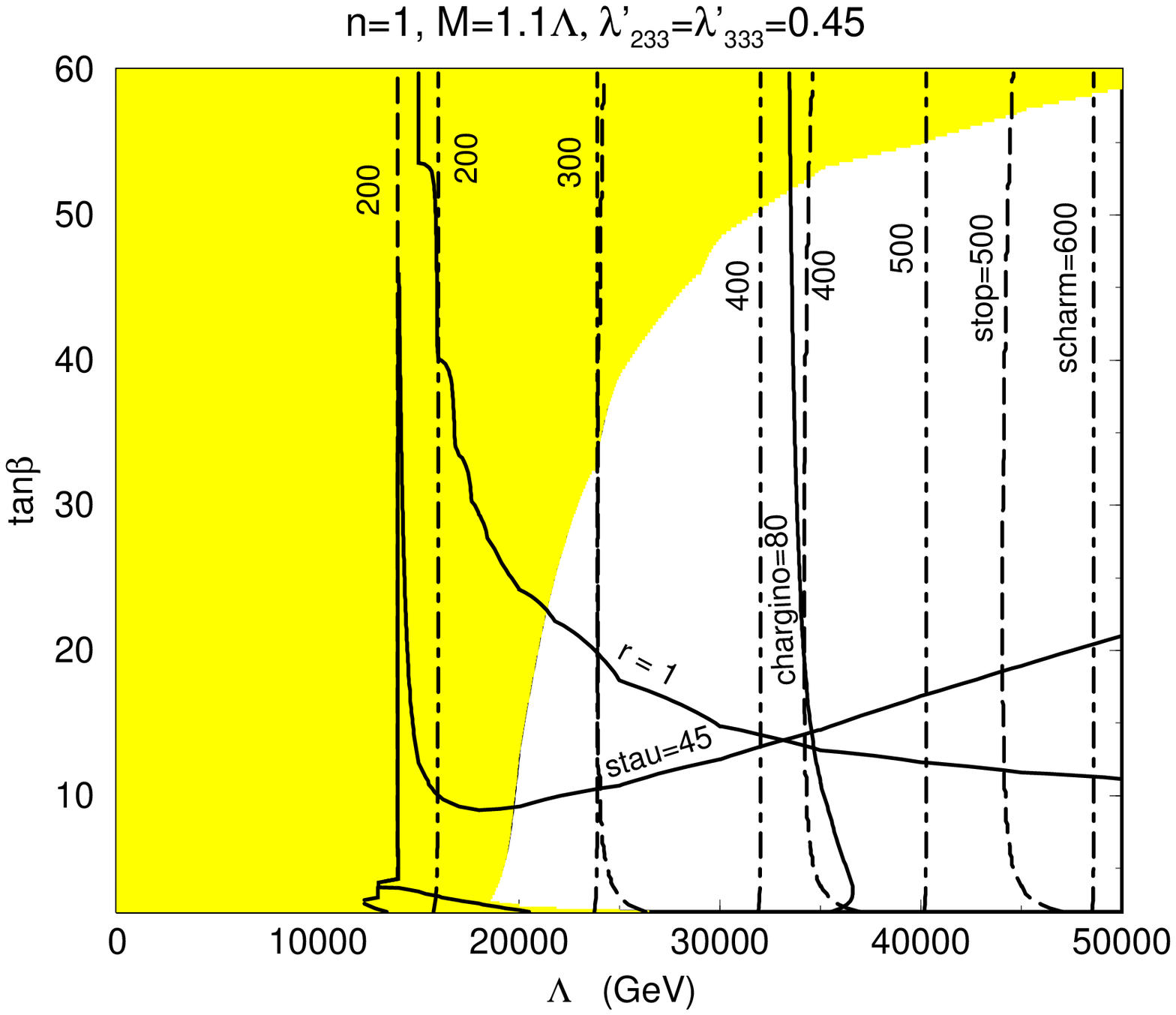}
\includegraphics[height=2.4in]{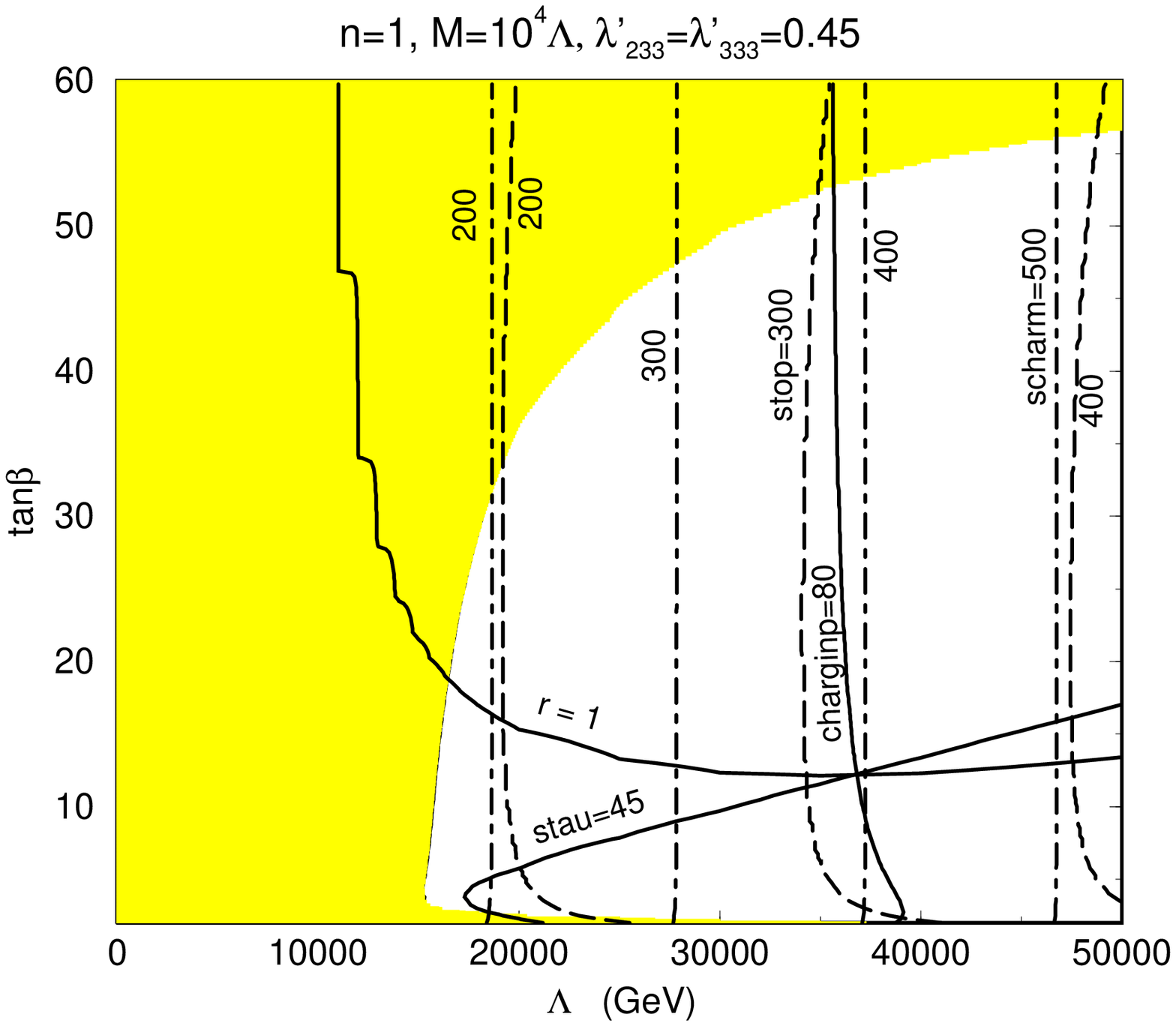}

\includegraphics[height=2.4in]{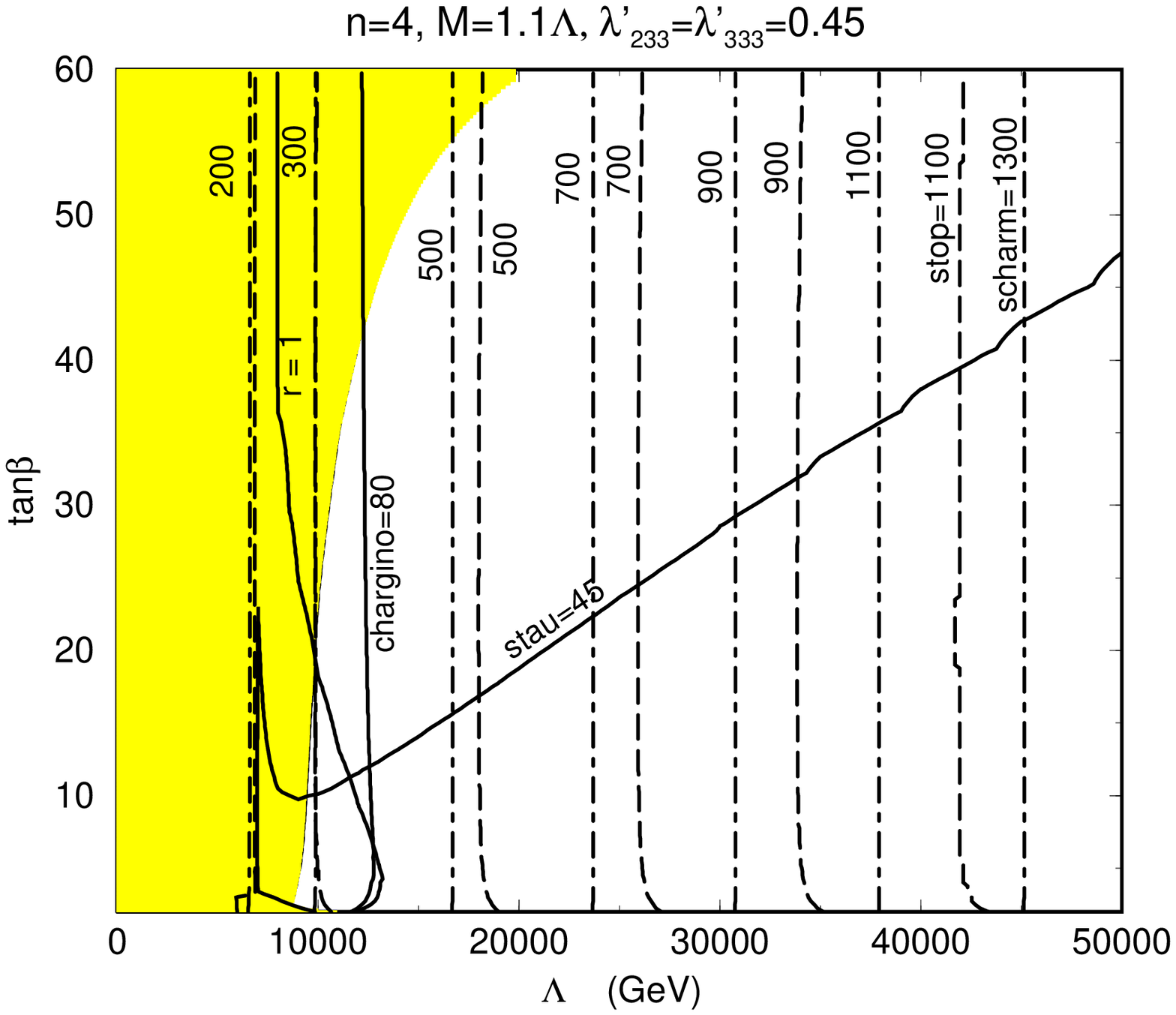}
\includegraphics[height=2.4in]{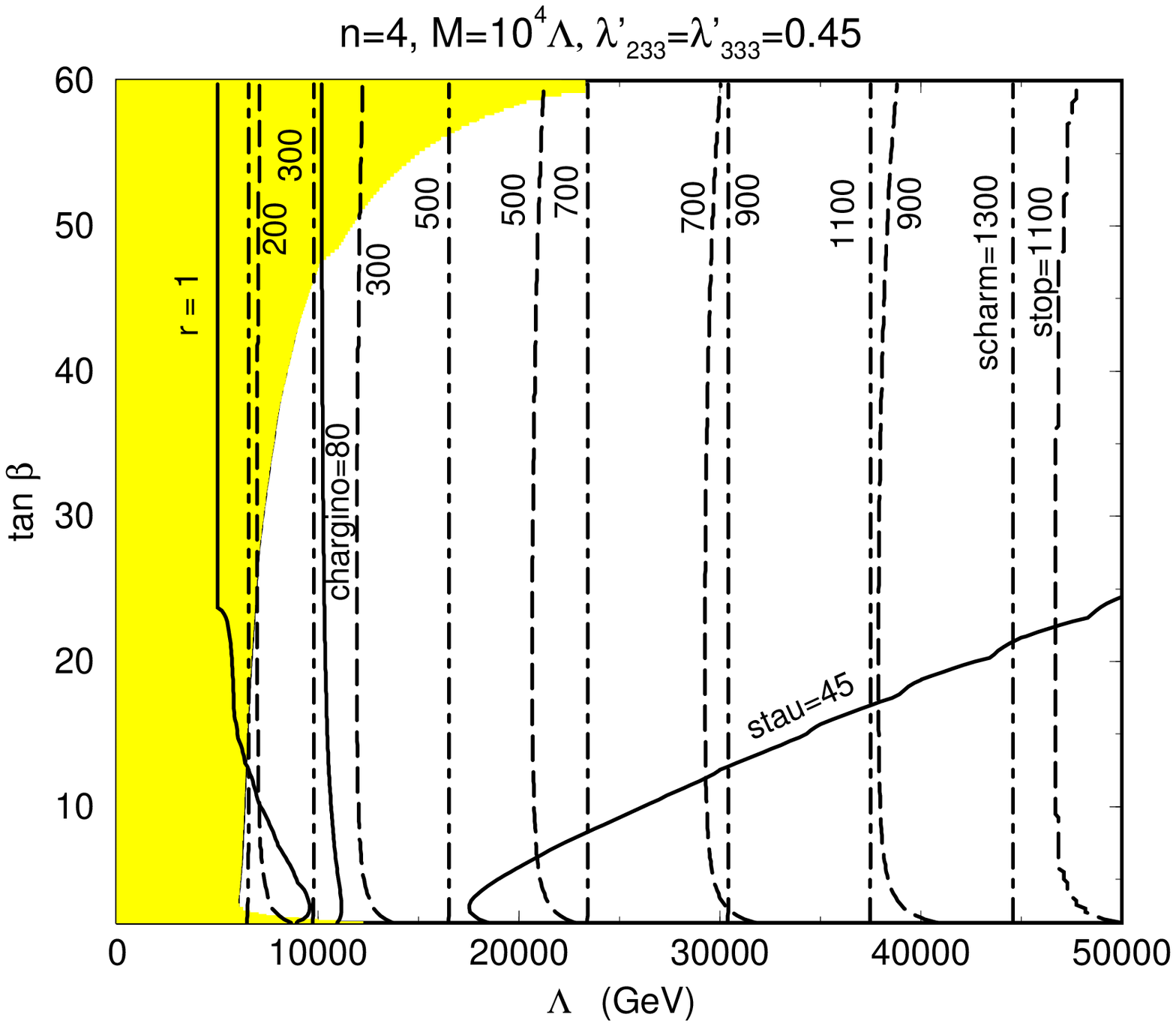}
\end{center}
\caption{\small  \label{fig1}
Contour plots of scalar top (dashed) and scalar charm (dot-dashed)
masses in the plane of $\Lambda$ vs $\tan \beta$. The shaded region is
excluded by the radiative electroweak-symmetry breaking and $M_{h^0}<60$
GeV.  The lightest chargino mass $M_{\tilde{\chi}^+_1}=80$ GeV, the
scalar tau $M_{\tilde{\tau}_1}=45$ GeV, and the ratio $r \equiv
M_{\tilde{\tau}_1} / M_{\tilde{\chi}^0_1} = 1$ are also plotted. The
region excluded by the chargino and the scalar tau masses are:
(i) when the neutralino is the NLSP (i.e. below or to the left of the
curve $r=1$) the region to the left of the curve $M_{\tilde{\chi}^+_1}= 80$
GeV is excluded; (ii) when the scalar tau is the NLSP (i.e. above or
to the right of the curve $r=1$) the region above or to the left of
the curve $M_{\tilde{\tau}_1}= 45$ GeV is excluded. }
\end{figure}

\begin{figure}[ht]
\leavevmode
\begin{center}
\includegraphics[height=3in]{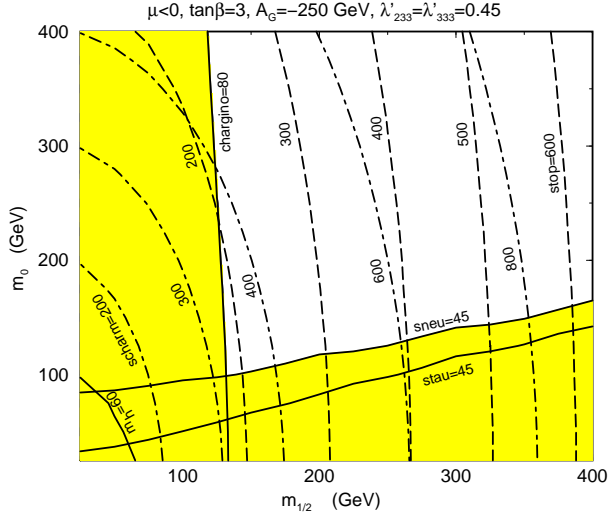}
\end{center}
\caption{\small \label{fig2}
Contour plots of the scalar top $\tilde{t}_1$ (dashed) and the scalar
charm $\tilde{c}_L$ (dot-dashed) in the plane of $m_{1/2}$ vs $m_0$ in
supergravity models.  The value of $A_G$ at the GUT scale is chosen as
$A_G=-250$ GeV. In addition, we use  
$\tan\beta=3$, and $\lambda'_{233}=\lambda'_{333}=
0.45$.  The shaded region is excluded by the constraints: $m_{h^0}<60$
GeV, $M_{\tilde{\chi}^+_1}<80$ GeV, $M_{\tilde{\tau}_1}<45$ GeV, and
$M_{\tilde{\nu}_{\tau_L}}<45$ GeV.
}
\end{figure}

Figures \ref{fig1} and \ref{fig2} show our main result.  From Fig. \ref{fig1}
we conclude that it is very difficult, if not impossible, to generate a
squark, either $\tilde{c}_L$ or $\tilde{t}_1$, of mass 200 GeV in
minimal gauge-mediated models for $n=1-4$ and $M/\Lambda=1.1-10^4$.
In Fig. \ref{fig1}, the shaded regions are excluded by the lower bound
on the lighter neutral Higgs boson, $m_{h^0}<60$ GeV, which
already covers the radiative electroweak-symmetry breaking.
Contours of
squark masses for $M_{\tilde{c}_L}$ (dot-dashed) and $M_{\tilde{t}_1}$
(dashed) are shown.  It is clear that a squark mass of 200 GeV is
almost impossible in all 4 cases shown, $n=1,4$ and $M/\Lambda=
1.1,10^4$.  Up to this point, we have not used any mass limits on the
SUSY particles.  Constraints on SUSY particle masses depend on whether the
neutralino or the scalar tau is the NLSP (the gravitino is the LSP and
$\tilde{\nu}_{\tau_L}$ is heavier than $\tilde{\tau}_1$ in most of the 
parameter space.)
In Fig. \ref{fig1}, we show the contour of $r\equiv
M_{\tilde{\tau}_1}/M_{\tilde{\chi}^0_1} =1$; above or to the right of
this contour $\tilde{\tau}_1$ is the NLSP, otherwise
$\tilde{\chi}^0_1$ is the NLSP.
When $\tilde{\tau}_1$ is the NLSP we use the constraint
$M_{\tilde{\tau}_1} > 45$ GeV; when $\tilde{\chi}^0_1$ is the NLSP we
use the chargino-mass constraint $M_{\tilde{\chi}^+_1} > 80$ GeV
\cite{aleph}.  In RPV theories, 
the chargino can decay into jets or multi-jets plus leptons or missing energy,
%
%
which ALEPH has searched for and put a bound on
$M_{\tilde{\chi}^+_1} > 83-85$ GeV \cite{aleph} 
(we use a conservative value of 80 GeV); when
$\tilde{\tau}_1$ is the NLSP there is no published limit, but we argue
that $\tilde{\tau}_1$ decays into jets, $\tilde{\tau}_1 \to qq'$, via
RPV couplings and should have been copiously seen in LEP1 if
$M_{\tilde{\tau}_1} <45$ GeV.
{}From Fig. \ref{fig1} this chargino or scalar tau mass
constraint can easily exclude squark masses up to 300 GeV for $n=1-4$
and $M/\Lambda=1.1-10^4$.

We have used large RPV couplings
($\lambda'_{333}=\lambda'_{233}$=0.45) in the figures.  Other values
of $\lambda'$ give similar results.  
We have repeated our RGE analysis with much
smaller $\lambda'_{233}=\lambda'_{333}=0.01$.  We found that all
squark, chargino, and the neutral Higgs boson masses are very
similar, but the scalar tau mass changes substantially. 
This does not affect our conclusion that 200 GeV squarks are not allowed 
in the parameter space constrained by the above requirements in 
gauge-mediated models.

If the value of the trilinear coupling $A$ is chosen to be large
numerically and to be of the same sign as $\mu$, the off-diagonal
matrix elements in the stop mass matrix will be very large, which could
lower the lighter stop mass.  We found that for $n=1$ and 
$M=10^4 \Lambda$ with $A_M= -500$
GeV we can produce a $\tilde{t}_1$ of mass 200 GeV in the parameter
space allowed by the constraints discussed above. Unfortunately, there
is no compelling reason for such a large $A_M$ in gauge-mediated
models.

In supergravity models, the value of $A_G$ at the GUT scale need not
be small.  For example, in the dilaton model $A_G=-m_{1/2}$ is naturally
negative and of order of a few hundred GeV.  We show in
Fig. \ref{fig2} the contours of $M_{\tilde{t}_1}$ and
$M_{\tilde{c}_L}$ in the supergravity model with $\tan\beta=3,\;
\mu<0$, and $A_G=-250$ GeV.  
The LSP can be $\tilde{\chi}^0_1$, $\tilde{\tau}_1$, or $\tilde{\nu}_{\tau_L}$.
The shaded region is excluded by $M_{\tilde{\chi}^+_1}<80$ GeV, 
$M_{\tilde{\tau}_1}<45$ GeV, and $M_{\tilde{\nu}_{\tau_L}}<45$ GeV.
Results for large $\tan\beta$ are
similar.  The effect of the RPV couplings on the squark masses
is again small but large on the scalar tau mass.  
Thus, it is possible to generate a stop of mass 200 GeV in the
supergravity models. However, it is unlikely to generate a light
scalar charm because the mixing is much smaller.


The gravitino is the stable LSP in gauge-mediated models even in
the presence of $R$-parity violation.  However, 
RPV couplings nearly forbid the decay of SUSY 
particles into gravitinos because 
the RPV couplings are much stronger than the gravitino coupling strength,
which is suppressed by the SUSY-breaking scale.
The lightest neutralino decays into a lepton
plus two jets, $\ell qq'$ or $\nu qq'$, via the RPV
couplings (the quark-squark mode is less favored because squarks are heavy.)
Unlike the gauge-mediated models with $R$-parity conservation, there
are no hard photons in the final state of SUSY particle decays in 
gauge-mediated
models with $R$-parity violation.  In supergravity-motivated
models with $R$-parity violation, the lightest neutralino decays with
similar signatures.  Consequently, these two different
SUSY-breaking scenarios cannot be distinguished by the decay
patterns of the SUSY particles if $R$-parity is
violated. Nevertheless, evidence of squarks ($\tilde{c}_L$ or
$\tilde{t}_1$) of mass around 200 GeV can distinguish rather cleanly
between gauge-mediated and supergravity-motivated models.

\section*{\bf Acknowledgments} We thank S. Nandi and Xerxes Tata for 
useful discussions.
This research was supported in part by the U.S. DOE under
Grants No.~DE-FG03-93ER40757 and DE-FG03-96ER40969


\end{document}